\documentstyle[preprint,aps,eqsecnum]{revtex}
\tightenlines                                                                                                                                                                
\newcommand{\be}{\begin{equation}}
\newcommand{\ee}{\end{equation}}
\newcommand{\bea}{\begin{eqnarray}}
\newcommand{\eea}{\end{eqnarray}}
\newcommand{\ba}{\begin{array}}
\newcommand{\ea}{\end{array}}
\newcommand{\vp}{\varphi}

\newcommand{\Th}{\Theta}
\newcommand{\tht}{\theta}

\newcommand{\la}{\lambda}

\newcommand{\de}{\delta}

\newcommand{\pa}{\partial}

\newcommand{\no}{\nonumber}

\newcommand{\Om}{\Omega}
\newcommand{\om}{\omega}

\newcommand{\sres}{\mbox{sres}}
\newcommand{\res}{\mbox{res}}
\newcommand{\Str}{\mbox{Str}}

\newcommand{\lb}{\label}

\newcommand{\hL}{\hat{L}}
\newcommand{\tL}{\tilde{L}}
\newcommand{\tH}{\tilde{H}}
\newcommand{\th}{\tilde{h}}
\newcommand{\hd}{\hat{d}}
\newcommand{\td}{\tilde{d}}

\begin{document}

\title{Hamiltonian Structures of Generalized  Manin-Radul Super KdV\\
and Constrained Super KP Hierarchies}

\author{Ming-Hsien Tu$^1$ and Jiin-Chang Shaw$^2$}
\address{
$^1$ Department of Mathematics, 
National Chung Cheng University, \\
Mingshiung, Chiayi 621, Taiwan\\
$^2$ Department of Applied Mathematics, 
National Chiao Tung University, \\
Hsinchu 300, Taiwan
}
\date{\today}
\maketitle

\begin{abstract}
A study of Hamiltonian structures associated with supersymmetric
Lax operators is presented. Following a constructive approach, 
the Hamiltonian structures of Inami-Kanno super KdV hierarchy 
and constrained modified super KP hierarchy are investigated from 
the reduced supersymmetric Gelfand-Dickey brackets. By applying 
a gauge transformation on the Hamiltonian structures associated with 
these two nonstandard super Lax hierarchies, we obtain the Hamiltonian 
structures of generalized Manin-Radul super KdV and constrained 
super KP hierarchies. We also work out a few examples and compare 
them with the known results.
\end{abstract}

\newpage
\section{Introduction}
In the past decade and more, the supersymmetric integrable
systems have received much attention in 
the literature (for recent reviews, see \cite{Kuper,S,Aratyn} and references therein), 
especially in the explorations of the relationship
 to the supersymmetric conformal field theories and string theories. 
On the one hand, in superconformal/superstring theories, correlation functions  
are governed by supersymmetric extensions of the Korteweg-de Vries (KdV)  
(or Kadomtsev-Petviashvili (KP)) systems. On the other hand, the knowledge of
super KdV/KP systems have motivated people to study non-perturbative 
properties of superstrings. These super integrable systems share many 
features in common: they have supersymmetric Lax representations, 
infinitely many conserved quantities and 
soliton solutions etc. Furthermore, it is a common belief that they 
also possess bi-Hamiltonian structures which define the dynamical
flows on the corresponding  Poisson supermanifolds. In particular, for the super KdV-type
systems, the Poisson brackets relative to their associated second Hamiltonian structures 
provide extended superconformal algebras ($W$-superalgebras) whose quantum versions
 serve as the highest weight representations of some infinite-dimensional  symmetries 
in string theories.

The main purpose of this paper is to construct the Hamiltonian structures of the
generalized Manin-Radul super KdV (MR sKdV) and constrained super KP
(csKP) hierarchies (for the definitions  of these hierarchies, see Sec. IV.) 
using the method of gauge transformtion. Although the Hamiltonian structures
for the simplest cases have been obtained in \cite{OP,AD}, however, to our knowledge, 
those for the general cases are still unexplored. Our motivation
comes from the fact that, for two gauge-equivalent integrable systems,
the  gauge transformation between them transforms  not only  the Lax formulations
but also the Hamiltonian structures of the corresponding hierarchies.
Hence the preparation of suitable super integrable systems which are gauge equivalent
to the generalized MR sKdV and csKP hierarchies is the key in this approach.
Our stretagy is the following: First, for an odd order super Lax operator $\hL$, 
we consider its associated supersymmetric Gelfand-Dickey (GD) bracket \cite{FR1} 
defined by the Hamiltonian map $J$. 
We then consider an usual reduction which modifies the Hamiltonian map $J$ to
$J_c$. Second,  we construct out two nonstandard super hierarchies from $(\hL, J_c)$
which have super Lax operators defined by $K_A=\hL D$ and $K_B=D^{-1}\hL$,
respectively. The former is referred to the Inami-Kanno sKdV (IK sKdV) hierarchy \cite{IK} 
whereas the latter to the constrained modified sKP (cmsKP) hierarchy \cite{Liu2,DG1,DG2}.
The Hamiltonian structures associated with $K_i$ can also be constructed
from $J_c$ and are denoted by $\Om^{(i)}$ ($i=A,B$). Finally, we perform a gauge
transformation on the systems $(K_i, \Om^{(i)})$ and denote the resulting systems 
by $(\tL_i, \Th^{(i)})$ which describe the Lax operators and the Hamiltonian structures of 
the generalized MR sKdV and csKP hierarchies.

In summary, we shall follow the following steps to achieve the goal
\be
(\hL, J_c)\to (K_i, \Om^{(i)}) \to (\tL_i, \Th^{(i)})
\ee
It will be shown below that each step described above automatically guarantees
the requirement that the associated Hamiltonian structures should obey the 
super Jacobi identity.

We organize this paper as follows: In Sec. II, we recall some basic facts concerning
super pseudodifferential operators (SPDOs). 
We then introduce the second supersymmetric
GD bracket and its reduction from a Miura transformation viewpoint. In Sec. III,
the IK sKdV and the cmsKP hierarchies are defined. 
We give a detailed construction of their associated Hamiltonian structures from the reduced 
supersymmetric GD bracket. We find that, up to a sign, the Poisson brackets defined 
by their corresponding Lax operators have the same form. 
In Sec. IV, we define the generalized MR sKdV and csKP hierarchies
by applying a gauge transformation to the IK sKdV and cmsKP hierarchies, 
respectively. We also show that this gauge transformation
enables us to obtain the Hamiltonian structures associated with the generalized MR sKdV and 
csKP hierarchies. In Sec. V, we give several examples to compare them with the
known results. We present our concluding remarks in Sec. VI.

\section{Supersymmetric Gelfand-Dickey brackets}
To begin with, we consider the supersymmetric Lax
operator of the form
\be
L=D^n+U_{n-1}D^{n-1}+\cdots+U_0
\lb{lax1}
\ee
where the supercovariant derivative $D\equiv\pa_{\tht}+\tht\pa$ $(\pa\equiv \pa/\pa_x)$ 
satisfies $D^2=\pa$, $\tht$ is the Grassmann variable ($\tht^2=0$) 
which together with the even variable $x\equiv t_1$ defines 
the $(1|1)$ superspace with coordinate $(x,\tht)$. 
The coefficients $U_i$ are superfields that depend on
the variables $\tht$, $t_i$ and can be represented 
by $U_i=u_i(t)+\tht v_i(t)$. 
The parity of a superfield $U$ is denoted by $|U|$ which is 
zero for $U$ being even and one for $U$ being odd.
Since $L$ is assumed to be homogeneous 
under $Z_2$-grading, thus $|U_i|=n+i$ (mod 2). 
We will introduce the Poisson bracket associated with
$L$ on functionals of the form:
\be
F(U)=\int_B f(U)
\ee
where $f(U)$ is a homogeneous differential polynomial 
of $U_i$ and $\int_B\equiv \int dxd\tht$ is the Berezin integral 
such that if $f(U)=a(u,v)+\tht b(u,v)$ then $\int_Bf(U)=\int b$. 
The supercovariant derivative $D$ satisfies the 
supersymmetric version of the Leibniz rule \cite{Manin}:
\be
D^iU=\sum_{k=0}^{\infty}(-1)^{|U|(i-k)} {i\brack k}U^{[k]}D^{i-k}
\label{leib}
\ee
where $U^{[k]}\equiv (D^kU)$ and the super-binomial 
coefficients ${i\brack k}$ are defined by
\be
{i\brack k}=
\left\{
\ba{l}
{[i/2]\choose [k/2]} \qquad \mbox{for $ 0\leq k\leq i$ 
and $(i,k)\ne (0,1)$ mod 2}\\
(-1)^{[k/2]}{-i+k-1\brack k}\qquad \mbox{for $i<0$}\\
0\qquad \mbox{otherwise}
\ea
\right.
\ee
For a given SPDO $P=\sum p_iD^i$ it is convenient to seperate $P$ into
differential part $P_+=\sum_{i\geq 0}p_iD^i$ and integral part
$P_-=\sum_{i\leq -1}p_iD^i$. In particular, we can define its super-residue as 
$\sres P=p_{-1}$ and its supertrace as $\Str P=\int_B\sres P$. 
It can be shown that, for any two SPDOs $P$ and $Q$,  $\Str [P,Q]=0$ 
for $[P,Q]\equiv PQ-(-1)^{|P||Q|}QP$ and hence $\Str PQ=(-1)^{|P||Q|}\Str QP$.
Given a functional $F(U)=\int_Bf(U)$, we define 
its gradient as
\be 
d_LF=\sum_{k=0}^{n-1}(-1)^kD^{-k-1}\frac{\de f}{\de U_k}
\lb{grad}
\ee
and its variation as
\be
\de F=(-1)^{|F|+|L|+1}\Str(\de Ld_LF)
\ee
where the variational derivative is defined by
\be
\frac{\de f}{\de U_k}=\sum_{i=0}^{\infty}(-1)^{|U_k|i+i(i+1)/2}
\left(\frac{\pa f}{\pa U_k^{[i]}}\right)^{[i]}
\lb{vari}
\ee
The supersymmetric second GD bracket associated with $L$ 
is given by \cite{FR1,FR2,FR3}
\be
\{F,G\}(L)=(-1)^{|F|+|G|+|L|+1}\Str[J(d_LF)d_LG]
\lb{sgd}
\ee
where the Hamiltonian map $J$ is defined by
\be
J(X)=(LX)_+L-L(XL)_+
\lb{gdhm}
\ee
where $X=\sum_kX_kD^k$. It has been shown \cite{FR1,FR3} 
that (\ref{sgd}) indeed defines a Hamiltonian structure, 
namely it is antisymmetric and satisfies the super Jacobi identity.

If we factorize $L=(D-\Phi_n)(D-\Phi_{n-1})\cdots(D-\Phi_1)$ which
defines a supersymmetric Miura transformation between the 
coefficient functions $U_i$ and the Miura fields $\Phi_i$,
then the second GD bracket (\ref{sgd}) becomes
\be
\{F,G\}(L)=\int_B\sum_{i=1}^n(-1)^i(D\frac{\de f}
{\de \Phi_i})\frac{\de g}{\de \Phi_i}
\lb{fact1}
\ee
which implies that the fundamental brackets of the Miura fields
$\Phi_i$ are given by \cite{FR1,FR2}
\be
\{\Phi_i(X),\Phi_j(Y)\}=(-1)^i\de_{ij}D\de(X-Y)
\lb{free}
\ee
where $X=(x,\tht)$, $Y=(y, \om)$ and $\de(X-Y)\equiv\de(x-y)(\tht-\om)$. 
This result is what we called the supersymmetric Kupershmidt-Wilson theorem.
Eq.(\ref{free}) enables us to write down the fundamental 
brackets of $U_k$ through the super Miura transformation. 

Next let us consider the case when the constraint $U_{n-1}=0$ is imposed in (\ref{lax1}).
It can be easily shown that such constraint for odd $n$ is second class which 
will modify the Hamiltonian structure $J$. 
On the other hand, for even $n$, the constraint is first class and hence the 
induced Poisson brackets can not be well defined .
Therefore for the odd order operator
$\hat{L}=D^{2k+1}+U_{2k-1}D^{2k-1}+\cdots+U_0$, we shall consider the factorization 
$\hat{L}=(D-\Phi_{2k+1})(D-\Phi_{2k})\cdots(D-\Phi_1)$. Then the modified
Poisson bracket defined by $\hat{L}$ becomes
\be
\{F,G\}_c=(-1)^{|F|+|G|}\Str(J_c(\hat{d}F)\hat{d}G)
\lb{rgdb}
\ee
where $\hat{d}F\equiv d_{\hL}F=
\sum_{i=0}^{2k-1}(-1)^iD^{-i-1}\frac{\de f}{\de U_i}$ and
\be
J_c(\hat{d}F)=J(\hat{d}F)+[\hat{L}, \int^x D\sres[\hat{L}, \hat{d}F]].
\ee
We remark that the second term is called the third GD structure which
is compatible with the second structure. 
Eq.(\ref{rgdb}) yields that the modified Poisson brackets
for the Miura fields $\Phi_i$ are given by
\be
\{\Phi_i(X),\Phi_j(Y)\}_c=[1+(-1)^i\de_{ij}]D\de (X-Y)
\lb{cfree}
\ee
which provide the free-field realizations of classical $W$-superalgebras
associated with the odd order Lax operator $\hL$\cite{FR2,HN,Huang}. Besides the usual
reduction described above, there are other reductions which have been
discussed in \cite{FR3,FMR}.
Since the first Hamiltonian structure can be obtained from the second
Hamiltonian structure by replacing $L$ by $L+\la$, where $\la$ is called
the spectral parameter, we shall focus only on the second structure.

\section{Two nonstandard Super Lax hierarchies}

There are several super integrable hierarchies whose Lax operators are related to the 
modifications  or reductions of the supersymmetric  Lax operator (\ref{lax1}) 
in the literature.  Here, for our purpose, we consider the following two Lax systems:
\be
\frac{d K_i}{d t_k}=[(K_i^{k/n})_{\geq 1}, K_i]\qquad (i=A,B)
\lb{eqk1}
\ee
with the Lax operators $K_i$ defined by
\bea
K_A&=&D^{2n}+V_{2n-2}D^{2n-2}+\cdots+V_1D
\lb{laxk1}\\
K_B&=&D^{2n}+V_{2n-2}D^{2n-2}+\cdots+V_0+D^{-1}V_{-1}
\lb{laxk2}
\eea
The Lax equation for $K_A$ is referred to the  IK sKdV hierarchy \cite{IK}.
The simplest example in this case is just the Laberge-Mathieu 
super KdV (LM sKdV) hierarchy ($n=2$) which was
constructed from a $N=2$ sKdV hierarchy\cite{Lab}.
On the other hand, the Lax equation for $K_B$ is the generalization
of the super two-boson hierarchy (sTB) ($n=1$) \cite{BD}, which we call
the cmsKP hierarchy.
In particular, from (\ref{eqk1}) it is easy to show that the coefficient 
function $V_{-1}$ obeys the evolution equation
\be
\frac{d V_{-1}}{d t_k}=-((K_B^{k/n})^*_{\geq 1}V_{-1}),
\lb{v-1}
\ee
which implies that $V_{-1}$ is an adjoint eigenfunction 
associated with the Lax operator $K_B$. 

In general the second Poisson brackets
associated with the Lax operators $K_i$ can be written as 
\be
\{F, G\}^{(i)}(K_i)=(-1)^{|F|+|G|+1}\Str(\Om^{(i)}(d_iF)d_iG)
\lb{pok1}
\ee
where $d_iF\equiv d_{K_i}F$ and the Hamiltonian maps 
$\Om^{(i)}$ are defined by 
\bea
\Om^{(A)}(d_AF)&=&(K_Ad_AF)_+K_A-K_A(d_AFK_A)_++[K_A, (d_AFK_A)_0]\no\\
& &+(-1)^{|F|}[\int^x D\sres[d_AF, K_A], K_A]+
(-1)^{|F|}K_AD^{-1}\sres[d_AF, K_A]
\lb{hmk1}\\
\Om^{(B)}(d_BF)&=&(K_Bd_BF)_+K_B-K_B(d_BFK_B)_++[K_B, (K_Bd_BF)_0]\no\\
& &+(-1)^{|F|}[K_B, \int^x D\sres[d_BF, K_B]]+
(-1)^{|F|}D^{-1}\sres[d_BF, K_B]K_B
\lb{hmk2}
\eea
Notice that the map $\Om^{(A)}$, in operator form,
 is similar to but different from $\Om^{(B)}$. 
Instead of giving $\Om^{(i)}$ by other methods \cite{Liu2,DG1,DG2,DP},
we will follow a constructive approach, analogous to that of the 
 supersymmetric GD structure \cite{FR1}, to verify the 
Hamiltonian maps $\Om^{(i)}$ from a supersymmetric 
Miura transformation point of view.
To show that the maps $\Om^{(i)}$ are indeed Hamiltonian 
we have to check that the Poisson brackets defined in (\ref{pok1})   
are antisymmetric and obey the super Jacobi identity. 
For antisymmetry, by direct computation, it can be easily shown that
\be
\{F,G\}^{(i)}=-(-1)^{|F||G|}\{G,F\}^{(i)}
\ee
For the super Jacobi identity, instead of direct computation,
we rewrite the Lax operator $K_i$ as 
\be
K_A=\hL_AD,\qquad K_B=D^{-1}\hL_B
\lb{veck2}
\ee
where $\hL_A$ and $\hL_B$ are super differential operators with order 
$2n-1$ and $2n+1$, respectively. 
Furthermore, from the relation 
\be
\de F=(-1)^{|F|+1}\Str(\de K_id_iF)=
(-1)^{|F|}\Str(\de \hL_i\hd_iF)
\ee
where $\hat{d}_i\equiv d_{\hL_i}$, we have
\be
\hd_AF=-Dd_AF,\qquad \hd_BF=(-1)^{|F|}d_BFD^{-1}
\lb{formk2}
\ee
Substituting (\ref{veck2}) and (\ref{formk2}) into (\ref{hmk1}) and (\ref{hmk2}) 
we find
\be
\Om^{(A)}(d_AF)=-J_c(\hd_AF)D,\qquad \Om^{(B)}(d_BF)=(-1)^{|F|}D^{-1}J_c(\hd_BF)
\ee
which imply that the Poisson brackets defined by $K_i$ can
be transformed to those defined by $\hL_i$  as follows
\be
\{F,G\}^{(i)}(K_i)=\eta_i\{F,G\}_c(\hL_i)
\lb{mpk1}
\ee
where $\eta_A=-1$ and $\eta_B=+1$.
Hence the super Jacobi identity associated with the maps $\Om^{(i)}$
is automatically satisfied due to the fact that the reduced supersymmetric 
GD brackets defined by $\hL_i$ admit  
Miura representations (\ref{cfree}).

Therefore the maps $\Om^{(i)}$ provide the Hamiltonian 
formulation for the Lax equations (\ref{eqk1}):
\be
\frac{d K_i}{dt_k}=\{H^{(i)}_k,K_i\}^{(i)}=\Om^{(i)}(d_iH^{(i)}_k)
\lb{hfk1}
\ee
where the Hamiltonian functionals $H_k^{(i)}$ are given by
\be
H^{(i)}_k=-\frac{n}{k}\Str(K_i^{k/n})
\lb{hk1}
\ee
Notice that the relative signs in the Hamiltonian maps $\Om^{(i)}$ 
 are crucial. It is this choice so that $\Om^{(i)}(d_iH_k^{(i)})$ are
differential operators of order less than $2n-2$ and Eq.(\ref{hfk1}) 
makes sense.

Before ending this section, two remarks are in order. 
First, we note that both Piosson brackets defined by $K_i$, up to a sign, are mapped to 
the same reduced supersymmetric GD bracket defined by $\hL_i$,  
which is different from the situation in the bosonic 
case where type-A is mapped to the {\it difference\/} 
of the second and the third GD structures \cite{Liu1} whereas 
type-B is the {\it sum} of the second and the third 
ones \cite{Liu1,ST}. Second, both Lax operators $K_A$ and $K_B$ 
can be factorized into  multiplicative forms, i.e.,
\bea
K_A&=&(D-\Phi_{2n-1})(D-\Phi_{2n-2})\cdots(D-\Phi_1)D\no\\
K_B&=&D^{-1}(D-\Phi_{2n+1})(D-\Phi_{2n})\cdots(D-\Phi_1)
\eea
where the Miura fields $\Phi_i$ obey the Poisson brackets
\be
\{\Phi_j(X),\Phi_k(Y)\}^{(i)}=\eta_i[1+(-1)^j\de_{jk}]D\de(X-Y)
\lb{rfree}
\ee

\section{generalized MR sKdV and constrained SKP hierarchies}

Having constructed the Hamiltonian structures of two nonstandard super
Lax  hierarchies in the previous section we are 
now ready to discuss gauge equivalences related to these two 
nonstandard hierarchies. 
Based on the fact that gauge transformations are
canonical transformations, we can use them to obtain new 
integrable Hamiltonian systems from the known ones. 
In the following, we will show that the second Hamiltonian structures 
of the generalized MR sKdV and csKP hierarchies are just
the ones which can be obtained in this way.

Let us perform the following gauge transformation
to the Lax operators $K_i$
\be
\tL_i=T^{-1}K_iT \qquad (i=A,B)
\label{gtl1}
\ee
where the gauge operator $T$ is defined by 
$T=\exp(-\int^x V_{2n-2}/n)$ and hence the next
leading term of $K_i$ can be gauged away.
The resulting differential operators $\tL_i$ are thus given by
\bea
\tL_A&=&D^{2n}+U_{2n-3}D^{2n-3}+\cdots+U_0
\lb{gtla}\no\\
\tL_B&=&D^{2n}+U_{2n-3}D^{2n-3}+\cdots+U_0+\phi D^{-1}\psi
\lb{gtlb}
\eea
where $\phi\equiv T^{-1}$ and $\psi\equiv V_{-1}T$.
It can be proved that $T^{-1}$ is an even eigenfunction associated with 
the operator $\tL_i$, i.e., 
$\pa T^{-1}/\pa t_k=((\tL_i^{k/n})_+T^{-1})_0$
and the nonstandard Lax equations in (\ref{eqk1}) are then
transformed to the standard ones 
\be
\frac{d \tL_i }{dt_k}=[(\tL_i^{k/n})_+, \tL_i]
\lb{eql1}
\ee
Therefore the gauge transformation (\ref{gtl1}) provides a connection
between $K_i$ and $\tL_i$ in the Lax formulation.  For $\tL_A$, the
Lax equation (\ref{eql1}) gives the  generalization of the MR sKdV hierarchy
($n=2$) which was originally constructed from the MR sKP hierarchy by reduction\cite{Manin}. 
On the other hand, the Lax equation (\ref{eql1}) for $\tL_B$ describes the 
csKP hierarchy  which contains the sAKNS hierarchy ($n=1$) \cite{AD,AR} 
as the simplest example. 
It can be easily shown that the Lax equation (\ref{eql1}) for $\tL_B$ is consistent with the 
following equations
\be
\frac{\pa \phi}{\pa t_k}=((\tL_B^{k/n})_+\phi)_0,\qquad
\frac{\pa \psi}{\pa t_k}=-((\tL_B^{k/n})_+^*\psi)_0
\ee
and thus $\phi$ and $\psi$ are even eigenfunction and odd adjoint 
eigenfunction of the csKP hierarchy, respectively. 

Moreover, since the hierarchy flows associated with $K_i$ have 
Hamiltonian descriptions, it is quite natural to ask whether we can use such 
gauge equivalence to obtain the second Hamiltonian structures of the generalized 
MR sKdV and csKP hierarchies. The answer is yes. 
To see this, consider an infinitesimal gauge transformation
$K_i\to K_i+Q$ where $Q$ is a homogeneous superdifferential 
operator of order at most $2n-2$. Then, in view of (\ref{gtl1}), we can read off the 
linearized map $T'$ and its transposed map $T'^{\dagger}$ as
\bea
T'&:& Q\to T^{-1}QT+\frac{1}{n}[\int^x q_{2n-2}, \tL_i]
\lb{limp}\\
T'^{\dagger} &:& P\to TPT^{-1}+\frac{(-1)^{|P|+1}}{n}
\int^x \sres[P, \tL_i]
\lb{tpmp}
\eea
where $P$ is an arbitrary SPDO, $q_{2n-2}\equiv \sres(QD^{-2n+1})$ 
and the adjoint of an operator $R$  is defined 
by $\Str(PRQ)=(-1)^{|R||P|}\Str(R^{\dagger}PQ)$.
Using $T'$ and $T'^{\dagger}$, a straightforward but tedius
calculation [see Appendix A] shows that
\bea
T'\Om^{(i)}T'^{\dagger}(P)
&=&(\tL_iP)_+\tL_i-
\tL_i(P\tL_i)_++\frac{1}{n}[\int^x \res [P, \tL_i], \tL_i]\no\\
&+&\frac{1}{n}[(\int^x \sres [P, \tL_i])D, \tL_i]-
\frac{2}{n^2}[\int^x((\int^{x'}\sres[P, \tL_i])U_{2n-3}), \tL_i]
\lb{hmmr}\\
&\equiv&\Th^{(i)}(P)\no
\eea
That means $\Th^{(A)}$ and $\Th^{(B)}$ , in terms of their own 
Lax operators,  have the same form.
Since $\Th^{(i)}$ are canonical equivalent to the Hamiltonian map $\Om^{(i)}$, 
the Poisson brackets defined by $\Th^{(i)}$ are also antisymmetric and obey the super
Jacobi identity. As a result, $\Th^{(A)}$ ($\Th^{(B)}$) can be defined as the 
Hamiltonian map of the generalized MR sKdV (csKP) hierarchy.
A further consistent check shows that $\Th^{(i)}$ map the Hamiltonian 
one-forms $\td_i\tH_k^{(i)}$ to (pseudo-) superdifferential operators of 
order at most $2n-3$. Now we can write down the Hamiltonian flows associated 
with the Lax operators $\tL_i$ as
\be
\frac{d \tL_i}{d t_k}=\{\tH^{(i)}_k, \tL_i\}=\Th^{(i)}(\td_i\tH^{(i)}_k)
\lb{hfl1}
\ee
where the Hamiltonian functionals, in view of (\ref{hk1}) and 
(\ref{gtl1}), are defined by
\be
\tH^{(i)}_k=-\frac{n}{k}\Str \tL_i^{k/n}
\lb{hl1}
\ee 
From the Hamiltonian flows (\ref{hfl1}) we can read off the 
Poisson brackets for the coefficient functions of $\tL_i$.

In fact, for $\tL_B$, we can express the associated Poisson brackets for
$U_i$, $\phi$ and $\psi$ more precisely. 
Let us rewrite $\tL_B=l+\phi D^{-1}\psi$  and  denote $H=\int_B h$ as one of the
Hamiltonian functionals $\tH_k^{(B)}$. 
Then the Hamiltonian one-form can be expressed as
\be
\td_BH=d_lH+X
\lb{forml2}
\ee
where $X$ is a superdifferential operator and 
\be
d_lH=\sum_{k=0}^{2n-3}(-1)^kD^{-k-1}\frac{\de h}{\de U_k}
\ee
Then from the relation
\bea
\de H&=&-\Str((\de l+\de \phi D^{-1}\psi+\phi D^{-1}\de \psi)
(d_lH+X))\no\\
&=&-\Str(\de ld_lH)+\int_B(\de \phi\frac{\de h}
{\de \phi}+\de \psi\frac{\de h}{\de \psi})
\eea
we have the following identifications
\be
\frac{\de h}{\de \phi}=(X^*\psi)_0,\qquad 
\frac{\de h}{\de \psi}=(X\phi)_0
\lb{idl2}
\ee
Inserting  (\ref{forml2}) with $X$ satisfying (\ref{idl2}) 
into the Hamiltonian map $\Th^{(B)}$ gives
\bea
\frac{d l}{d t}&=&(ld_lH)_+l-l(d_lHl)_++((ld_lH)_+\phi D^{-1}\psi)_+-
(\phi D^{-1}\psi(d_lHl)_+)_+\no\\
& &+(l\frac{\de h}{\de \psi}D^{-1}\psi)_+-(\phi D^{-1}\frac{\de h}{\de \phi}l)_+
+\frac{1}{n}[\int^x\res[\td_BH, \tL_B], l]-
\frac{2}{n}\phi\psi\int^x\sres[\td_BH, \tL_B]\no\\
& &+\frac{1}{n}[\int^x\sres[\td_BH, \tL_B], l]+
\frac{2}{n^2}[\int^x(U_{2n-3}\int^{x'}\sres[\td_BH, \tL_B]), l]\no\\
\frac{d \phi}{d t}&=& ((ld_lH)_+\phi)_0+(l\frac{\de h}{\de \psi})_0+
\phi[\int^x(D\psi\frac{\de h}{\de \psi})-\int^x(D\phi\frac{\de h}{\de \phi})]
+\frac{1}{n}\phi\int^x\res[\td_BH, \tL_B]\no\\
& &-\frac{1}{n}(D\phi)\int^x\sres[\td_BH, \tL_B]
+\frac{2}{n^2}\phi\int^x(U_{2n-3}\int^{x'}\sres[\td_BH, \tL_B])\no\\
\frac{d \psi}{d t}&=&-((l^*(d_lH)^*)_+\psi)_0-(l^*\frac{\de h}{\de \phi})_0+
\psi[\int^x(D\phi\frac{\de h}{\de \phi})-\int^x(D\psi\frac{\de h}{\de \psi})]
-\frac{1}{n}\psi\int^x\res[\td_BH, \tL_B]\no\\
& & +\frac{1}{n}(D\psi\int^x\sres[\td_BH, \tL_B])
-\frac{2}{n^2}\psi\int^x(U_{2n-3}\int^{x'}\sres[\td_BH, \tL_B])
\lb{hfl2}
\eea
where
\bea
\res[\td_BH, \tL_B]&=&\res[d_lH, l]+(D\psi)\frac{\de h}{\de \psi}-
\phi(D\frac{\de h}{\de \phi})-\sres(d_lH\phi\psi)-\phi(D(d_lH)^*\psi)\no\\
\sres[\td_BH, \tL_B]&=&\sres[d_lH, l]-\psi\frac{\de h}{\de \psi}+\phi\frac{\de h}{\de \phi}
\eea
Eq.(\ref{hfl2}) can be viewed as the supersymmetric generalization of the second
Hamiltonian structures of constrained KP hierarchy derived by Oevel and Strampp \cite{OS}

\section{Examples}
In this section we work out a number of examples to illustrate the 
previous results explicitly. We write down the Poisson brackets for
these systems according to the formulas given above and compare them
with the known results.

\subsection{Laberge-Mathieu super KdV hierarchy}
For $K_A=\pa^2+v_2\pa+v_1D$ the first equations in 
(\ref{eqk1}) are given by
\bea
\frac{d}{dt_0}
\left(
\ba{c}
v_1\\
v_2
\ea
\right)
&=&\left(
\ba{c}
v_{1x}\\
v_{2x}
\ea
\right)\no\\
\frac{d}{dt_1}
\left(
\ba{c}
v_1\\
v_2
\ea
\right)
&=&\frac{1}{4}
\left(
\ba{c}(v_{1xx}+3v_1(Dv_1)-\frac{3}{2}v_1v_2^2-3v_1v_{2x})_x\\
(v_{2xx}-\frac{1}{2}v_2^3+3v_1(Dv_2))_x
\ea
\right)
\lb{eqlm}
\eea
which represents the first equations of the LM sKdV hierarchy. 
The Hamiltonian formulation for these 
equations are given by (\ref{hfk1}) where the second Poisson 
structure can be obtained by substituting 
$d_AH^{(A)}_k=-D^{-2}\frac{\de h^{(A)}_k}{\de v_1}+D^{-3}\frac{\de h^{(A)}_k}
{\de v_2}$ into (\ref{hmk1}). We find
\be
\frac{d}{dt_k}
\left(
\ba{c}
v_1\\
v_2
\ea
\right)
=\left(
\ba{cc}
-2v_1\pa-v_{1x} & -\pa^2-v_2\pa+v_1D-(Dv_1)\\
\pa^2-v_2\pa+v_1D-v_{2x} & -2D^3+(Dv_2)-2v_1
\ea
\right)
\left(
\ba{c}
\frac{\de h^{(A)}_k}{\de v_1}\\
\frac{\de h^{(A)}_k}{\de v_2}
\ea
\right)
\lb{hflm}
\ee
where the first Hamiltonian functionals are 
given by
\bea
H^{(A)}_0&=&-2\Str K_A^{1/2}=-\int_B v_1\no\\
H^{(A)}_1&=&-\frac{2}{3}\Str K_A^{3/2}=
-\frac{3}{8}\int_B[\frac{1}{2}v_1v_2^2+v_1v_{2x}-v_1(Dv_1)]
\eea
To compare with the known result, we consider the change 
of variables as follows
\be
(v_1,v_2)\to (-(Du)-\tau, -2u)
\ee
then the Poisson structure in (\ref{hflm}) becomes
\be
\frac{1}{2}\left(
\ba{cc}
-D\pa+\tau & 2u\pa-(Du)D+2u_x\\
2u\pa-(Du)D+u_x & -D\pa^2+3\tau\pa+(D\tau)D+2\tau_x\ea\right)
\lb{polm}
\ee
which is just the form presented in \cite{MP1}

\subsection{Super two-boson hierarchy} 
For $K_B=\pa+v_0+D^{-1}v_{-1}$ the first Lax 
equations in (\ref{eqk1}) are given by
\bea
\frac{d}{dt_1}
\left(
\ba{c}
v_0\\
v_{-1}
\ea
\right)
&=&
\left(
\ba{c}
v_{0x}\\
v_{-1x}
\ea
\right)\no\\
\frac{d}{dt_2}
\left(
\ba{c}
v_0\\
v_{-1}
\ea
\right)
&=&
\left(
\ba{c}
v_{0xx}+2(Dv_{-1})_x+(v_0^2)_x\\
-v_{-1xx}+2(v_0v_{-1})_x
\ea
\right)
\lb{eqtb}
\eea
which represents the first equations of the sTB hierarchy.
The Hamiltonian  description for these equations are
given by (\ref{hfk1}) where the second Poisson structure can be 
obtained by substituting $d_BH^{(B)}_k=D^{-1}\frac{\de h^{(B)}_k}{\de v_0}
+\frac{\de h^{(B)}_k}{\de v_{-1}}$ into (\ref{hmk2}). It turns out that
\be
\frac{d}{dt_k}
\left(
\ba{c}
v_0\\
v_{-1}
\ea
\right)=
\left(
\ba{cc}
2D^3+(Dv_0)+2v_{-1} & \pa^2+v_0\pa+v_{-1}D+v_{0x}\\
-\pa^2+v_0\pa+v_{-1}D-(Dv_{-1}) & 2v_{-1}\pa+v_{-1x}
\ea
\right)
\left(
\ba{c}
\frac{\de h^{(B)}_k}{\de v_0}\\
\frac{\de h^{(B)}_k}{\de v_{-1}}
\ea
\right)
\lb{hftb}
\ee
where the first Hamiltonian functionals 
are given by
\bea
H^{(B)}_1&=&-\Str K_B=-\int_B v_{-1}\no\\
H^{(B)}_2&=&-\frac{1}{2}\Str K_B^2=\int_B
v_0v_{-1}
\eea
Eq.(\ref{hftb}) provides the second Hamiltonian formulation
of the sTB hierarchy.

If we make the following identification
\be
(v_0, v_{-1})\to (-(DJ_0), J_1)
\ee 
then the second Poisson structure in (\ref{hftb}) becomes
\be
\left(
\ba{cc}
2D+2D^{-1}J_1D^{-1}- D^{-1}J_{0x}D^{-1}-& 
-D^3+D(DJ_0)-D^{-1}J_1D\\
D^3+(DJ_0)D+DJ_1D^{-1} & J_1D^2+D^2J_1
\lb{potb}
\ea
\right)
\ee
which is the form of the second Poisson structure discussed in \cite{BD}.

\subsection{Manin-Radul  super KdV hierarchy}
For $\tL_A=\pa^2-\vp D+a$ the first Lax equations in (\ref{eql1})
are given by
\bea
\frac{d}{dt_0}
\left(
\ba{c}
a\\
\vp
\ea
\right)
&=&
\left(
\ba{c}
a_x\\
\vp_x
\ea
\right)\no\\
\frac{d}{dt_1}
\left(
\ba{c}
a\\
\vp
\ea
\right)
&=&\frac{1}{4}
\left(
\ba{c}
\vp_{xxx}-3(\vp(D\vp))_x+6(a\vp)_x\\
a_{xxx}-3(\vp(Da))_x+3(a^2)_x
\ea
\right)
\lb{eqmr}
\eea
which represents the first equations of the MR sKdV hierarchy.
The Hamiltonian formulation of these equations are given by
(\ref{hfl1}) in which the first Hamiltonian functionals are given by
\bea
\tH^{(A)}_0&=&-2\Str \tL_A^{1/2}=\int_B\vp\no\\
\tH^{(A)}_1&=&-\frac{2}{3}\Str \tL_A^{3/2}=
-\frac{1}{4}\int_B [\vp(D\vp)-2\vp a]
\eea
and  the second Poisson structure can be obtained
by substituting $\td_A\tH^{(A)}_k=D^{-1}\frac{\de \th^{(A)}_k}{\de a}+D^{-2}
\frac{\de \th^{(A)}_k}{\de \vp}$ into (\ref{hmmr}). It turns out that
\be
\frac{d}{dt_k}
\left(
\ba{c}
a\\
\vp
\ea
\right)=
\left(
\ba{cc}
P_{aa} & P_{a\vp}\\
P_{\vp a} & P_{\vp\vp}
\ea
\right)
\left(
\ba{c}
\frac{\de \th^{(A)}_k}{\de a}\\
\frac{\de \th^{(A)}_k}{\de \vp}
\ea
\right)
\lb{hfmr}
\ee
where the second Poisson matrix is given by
\bea
P_{aa}&=& \frac{1}{2}[D\pa^3-3\vp\pa^2+4aD\pa+(2(Da)-3\vp_x)\pa
+2a_xD+3\vp(D\vp)+(D^3a)-4a\vp-\vp_{xx}\no\\
& &+\vp D^{-1}(Da)-(Da)D^{-1}\vp-\vp D^{-1}\vp D^{-1}\vp
-\vp D^{-1}\vp_x+\vp_x D^{-1}\vp]\no\\
P_{a\vp}&=& \frac{1}{2}[\pa^3-2\vp D\pa+4a\pa
-\vp_xD+2a_x+\vp D^{-1}(D\vp)]\no\\
P_{\vp a}&=&\frac{1}{2}[\pa^3+2\vp D\pa+(4a-2(D\vp))\pa
+\vp_xD+2a_x-(D^3\vp)+(D\vp)D^{-1}\vp]\no\\
P_{\vp\vp}&=&\frac{1}{2}[4\vp\pa+2\vp_x]
\lb{pomr}
\eea
Eq.(\ref{hfmr}) provides the second Hamiltonian formulation of 
the MR sKdV hierarchy reported in \cite{OP}.

Starting from the Lax operator $K_A=\pa^2+v_2\pa+v_1D$ associated
with the LM sKdV hierarchy one can perform the gauge transformation
$T=\exp(-\int^x v_2/2)$ \cite{IK} on the Lax operator $K_A$ as follows
\bea
K_A\to \tL_A&=&e^{\int^x v_2/2}K_Ae^{-\int^x v_2/2}\no\\
&=&\pa^2+v_1D-(\frac{v_2^2}{4}+\frac{v_2'}{2}+\frac{v_1(D^{-1}v_2)}{2})
\eea
Then the Lax operator $\tL_A=\pa^2-\phi D+a$ associated with the 
MR sKdV hierarchy is related to the Lax operator $K_A$ as
\be
\phi=-v_1,\qquad a=-(\frac{v_2^2}{4}+\frac{v_2'}{2}+\frac{v_1(D^{-1}v_2)}{2})
\ee
which provides the gauge equivalence between the LM sKdV hierarchy (\ref{eqlm}) 
and the MR sKdV hierarchy (\ref{eqmr}). Moreover it has been shown \cite{MP1} that
the second Hamiltonian structure (\ref{polm}) of the LM sKdV hierarchy can be transformed
to the second Hamiltonian structure (\ref{pomr}) of the MR sKdV hierarchy via this gauge
transformation.

\subsection{Super AKNS hierarchy} 

For $\tL_B=\pa+\phi D^{-1}\psi$ the first equations in (\ref{eql1}) are 
given by
\bea
\frac{d}{dt_1}\left(
\ba{c}
\phi\\
\psi
\ea
\right)
&=&\left(
\ba{c}
\phi_x\\
\psi_x
\ea
\right)\no\\
\frac{d}{dt_2}\left(
\ba{c}
\phi\\
\psi
\ea
\right)&=&
\left(
\ba{c}
\phi_{xx}+2\phi(D\phi\psi)\\
-\psi_{xx}-2\psi(D\phi\psi)
\ea
\right)
\lb{eqskp}
\eea
which are the first equations in the sAKNS hierarchy.
Hamiltonian formulations for these equations are given 
by (\ref{hfl2}), where the first Hamiltonian functions are given by
\bea
\tH^{(B)}_1&=&-\Str \tL_B=\int_B\phi\psi \no\\
\tH^{(B)}_2&=&-\frac{1}{2}\Str \tL_B^2=\int_B \phi_x\psi
\eea
From (\ref{hfl2}) the Hamiltonian flow can be expressed as
\be
\frac{d}{dt_k}\left(
\ba{c}
\phi\\
\psi
\ea
\right)=\left(
\ba{cc}
P_{\phi\phi} & P_{\phi\psi}\\
P_{\psi\phi} & P_{\psi\psi}
\ea
\right)
\left(
\ba{c}
\frac{\de \th^{(B)}_k}{\de \phi}\\
\frac{\de \th^{(B)}_k}{\de \psi}
\ea
\right)
\lb{hfskp}
\ee
where the Poisson brackets are given by
\bea
P_{\phi\phi}&=&-\phi D^{-1}\phi-\phi D^{-2}\phi D-(D\phi)D^{-2}
\phi-2\phi D^{-2}\phi\psi D^{-2}\phi \no\\
P_{\phi\psi}&=& D^2+\phi D^{-1}\psi+\phi D^{-2}(D\psi)+
(D\phi)D^{-2}\psi+2\phi D^{-2}\phi\psi D^{-2}\psi\no\\
P_{\psi\phi}&=&D^2+\psi D^{-2}\phi D+(D\psi)D^{-2}\phi+
2\psi D^{-2}\phi\psi D^{-2}\phi\no\\
P_{\psi\psi}&=&-(D\psi)D^{-2}\psi-\psi D^{-2}(D\psi)-
2\psi D^{-2}\phi\psi D^{-2}\psi
\lb{poskp}
\eea
which is just the second Poisson structure obtained in \cite{AD}.
Eq.(\ref{hfskp}) provides the
second Hamiltonian formulation of the sAKNS hierarchy.

Starting from the Lax operator $K_B=\pa+v_0+D^{-1}v_{-1}$ associated
with the sTB hierarchy one can perform the gauge transformation
$T=\exp(-\int^x v_0)$ \cite{AR,ST2} to the Lax operator $K_B$ as follows
\bea
K_B\to \tL_B&=&e^{\int^x v_0}K_Be^{-\int^x v_0}\no\\
&=&\pa+e^{\int^x v_0} D^{-1}e^{-\int^x v_0}
\eea
Then the Lax operator $\tL_B=\pa+\phi D^{-1}\psi$ associated with the 
sAKNS hierarchy is related to the Lax operator $K_B$ as
\be
\phi=e^{\int^x v_0},\qquad a=v_{-1}e^{-\int^x v_0}
\ee
which provides the gauge equivalence between the sTB hierarchy (\ref{eqtb}) 
and the sAKNS hierarchy (\ref{eqskp}). Moreover it can be proved \cite{ST2} that
the second Hamiltonian structure (\ref{potb}) of the sTB hierarchy can be transformed
to the second Hamiltonian structure (\ref{poskp}) of the sAKNS hierarchy via this gauge
transformation.

\section{concluding remarks}
In this paper, we investigate the Hamiltonian structures associated
with several supersymmetric extensions of the KdV hierarchy.
Starting with the reduced super GD bracket, the Hamiltonian structures
of two nonstandard super KdV hierarchies can be constructed via
supersymmetric Miura transformations. We then perform a gauge
transformation on these two nonstandard Lax hierarchies to obtain
the Hamiltonian structures of the generalized MR sKdV hierarchy and constrained 
sKP hierarchy in a unified fasion. 
To compare the obtained Hamiltonian structures with the known 
results, we work out a few examples including the LM sKdV, sTB, MR sKdV  and  
sAKNS hierarchies.

Our approach on the gauge transformation relies on the algebra of super
pseudo-differential operators which provides an effective method to achieve the
goal. In fact, the gauge transformation (\ref{gtl1}) which maps $\Om^{(i)}$ to 
$\Th^{(i)}$ is by no mean unique.  There is another gauge transformation
triggered by $S=D^{-1}T$ \cite{ST2,ST1} which also brings $\Om^{(i)}$ to $\Th^{(i)}$.
Since the parity of $S$ is odd, the gauge equivalence of the Hamiltonian maps
given by (\ref{hmmr}) should be replaced by $S'\Om^{(i)}S'^{\dagger}=-\Th^{(i)}$
where the minus sign will be compensated by that induced
from the transformation of the Hamiltonians such that the hierarchy flows (\ref{hfk1})
are transformed to (\ref{hfl1}).

Finally, we would like to comment briefly on the algebraic structures 
associated with the Poisson brackets defined by the Hamiltonian maps 
$\Om^{(i)}$ and $\Th^{(i)}$. As we shown in Eq.(\ref{mpk1}),
the Poisson brackets defined by $\Om^{(i)}$ are encoded by the Poisson bracket 
defined by $J_c$. However, it has been shown \cite{FR2,Huang} that in the space 
of supersymmetric Lax operator of odd order, the reduced supersymmetric 
GD bracket (\ref{rgdb})  defines an infinite series of classical $N=2$ $W$-superalgebras 
which contain $N=2$ super Virasoro algebra as a subalgebra. 
Therefore, through the Miura transformation, the differential polynomials of
the coefficient functions $V_i$ of $K_i$ can be identified as the $N=2$ supermultiplets and 
Eq.(\ref{rfree}) provides the free-field realizations of the corresponding $W$-superalgebras.
On the other hand, for the MR sKdV and csKP hierarchies, the Poisson algebras
defined by $\Th^{(i)}$ are not quite clear so far, even for the simplest case. 
It seems difficult to construct the super Virasoro generator by covariantizing the supersymmetric 
Lax operator $\tL_i$ due to the fact that $U_{2n-1}=U_{2n-2}=0$. 
Therefore, to explore the algebraic structures associated with $\Th^{(i)}$, 
the decompositions of coefficient functions $U_i$ into primary fields remain 
to be worked out. Work in this direction is still in progress.

{\bf Acknowledgments\/}

We would like to thank W.J. Huang for a number of helpful discussions
and M.C. Chang for reading the manuscript.
This work is supported by the National Science Council
of Taiwan under Grant Nunbers NSC 88-2811-M-194-0003(MHT) and 
NSC 88-2112-M-009-008 (JCS).

\newpage

\appendix
\section{Proof for (4.7)}

To prove (\ref{hmmr}), let $P$ be an arbitrary super pseudo-differential operator, then
\be
T'\Om^{(A)} T'^{\dagger}P=T'Q
\ee
where
\bea
Q&\equiv& \Om^{(A)} T'^{\dagger}P\no\\
&=&(K_AT'^{\dagger}P)_+K_A-K_A(T'^{\dagger}PK_A)_++
[K_A, (T'^{\dagger}PK_A)_0]\no\\
& &+(-1)^{|P|}[\int^x D\sres[T'^{\dagger}P, K_A], K_A]+
(-1)^{|P|}K_AD^{-1}\sres[T'^{\dagger}P, K_A]
\eea
Using (\ref{tpmp}), each term in $Q$ can be calculated as follows
\bea
(1)&=&(TLPT^{-1})_+K_A+\frac{(-1)^{|P|+1}}{n}D(\int^x\sres[P, L])K_A \no\\
(2)&=&-K_A(TPLT^{-1})_++\frac{(-1)^{|P|}}{n}K_A(D\int^x\sres[P, L])
-\frac{1}{n}(\int^x\sres[P, L])D \no\\
(3)&=&[K_A, (TPLT^{-1})_0]+\frac{(-1)^{|P|+1}}{n}[K_A, (D\int^x\sres[P, L])] \no\\
(4)&=&(5)=0 \no
\eea
which imply that
\be
Q=(TLPT^{-1})_+K_A-K_A(TPLT^{-1})_++[K_A, (TPLT^{-1})_0]+
\frac{1}{n}[(\int^x\sres[P, L])D, K_A]
\lb{B}
\ee
and
\bea
\frac{1}{n}\int^x q_{2n-2}&=&
\frac{1}{n}\int^x\sres(QD^{-2n+1})\no\\
&=&(TPLT^{-1})_0+\frac{1}{n}\int^x\res(T[P, L]T^{-1})+
\frac{1}{n}\int^x[(\int^{x'}\sres[P, L])\frac{(DV_{2n-2})}{n}]\no\\
& &-\frac{2}{n^2}\int^x[(\int^{x'}\sres[P, L])V_{2n-3}]
\lb{b}
\eea
Substituting (\ref{B}) and (\ref{b}) into (\ref{limp}) we obtain the desired result (\ref{hmmr}).

Since the proof for $K_B$ is parallel to the above one, hence we omit it here.

\end{document}